\documentclass[letterpaper, 10pt, conference]{IEEEtran}      
\pdfoutput=1

\usepackage{epsfig}
\usepackage{times}
\usepackage{float}
\usepackage{afterpage}
\usepackage{amstext}
\usepackage{amssymb,bm}
\usepackage{latexsym}
\usepackage{color}
\usepackage{amsmath}
\usepackage{theorem}
\usepackage{stfloats}
\usepackage{pstricks}
\usepackage{subfigure}

\begin{document}

\newcommand{\gapFD}{2}
\newcommand{\gapHD}{4.82}

\title{The Capacity of the Gaussian Cooperative Two-user Multiple Access Channel to within a Constant Gap}

\author{%
\IEEEauthorblockN{Daniela Tuninetti\\
University of Illinois at Chicago,
Chicago, IL 60607, USA, 
Email: danielat@uic.edu}
}
\maketitle

\begin{abstract}
The capacity region of the cooperative two-user Multiple Access Channel (MAC) in Gaussian noise is determined to within a constant gap for both the Full-Duplex (FD) and Half-Duplex (HD) case. The main contributions are: (a) for both FD and HD: unilateral cooperation suffices to achieve capacity to within a constant gap where only the user with the strongest link to the destination needs to engage in cooperation, (b) for both FD and HD: backward joint decoding is not necessary to achieve capacity to within a constant gap, and (c) for HD: time sharing between the case where the two users do not cooperate and the case where the user with the strongest link to the destination acts as pure relay for the other user suffices to achieve capacity to within a constant gap. These findings  show that simple achievable strategies are approximately optimal for all channel parameters with interesting implications for practical cooperative schemes.
\end{abstract}


\section{Introduction}

\subsection{The General Memoryless Two-User Cooperative MAC}
A cooperative two-user Multiple Access Channel (Coop2MAC) is a three node network, where the sources are referred to as user/node~1 and user/node~2, respectively, and the destination as node~3. The channel is specified by two input alphabets $(\mathcal{X}_1,\mathcal{X}_2 )$, three output alphabets $(\mathcal{Y}_1,\mathcal{Y}_2,\mathcal{Y}_3)$, and a memoryless channel with transition probability $\mathbb{P}_{Y_1,Y_2,Y_3|X_1,X_2}$. 
User~$i\in\{1,2\}$ has a message $W_i$ uniformly distributed on $[1:2^{N R_i}]$ and independent of everything else for the destination, where $N\in\mathbb{N}$ denotes the codeword length and $R_i\in\mathbb{R}_+$ the transmission rate in bits per channel use. At time $t\in [1:N]$ user~$i\in\{1,2\}$ sends $X_{i,t}( W_i,Y_i^{t-1})$. At time $N$ the destination outputs the message estimates $\widehat{W}_1(Y_3^N)$ and $\widehat{W}_2(Y_3^N)$. The capacity region is the convex closure of all rate pairs $(R_1,R_2)$ such that $\mathbb{P}[ (\widehat{W}_1,\widehat{W}_2) \neq (W_1,W_2)] \rightarrow 0$ as $N \rightarrow +\infty.$ 
The best outer bound for the Coop2MAC is obtained as the intersection of the cut-set upper bound~\cite{elgamalmim} and the dependance balance bound of Hekstra and Willems~\cite{kekstrawillems}. The best achievable region is due to Willems et al.~\cite{willemsMACGF} and uses block-Markov coding and backward decoding. These bounds are known to coincide for the case of common output feedback, i.e., $Y_1=Y_2=Y_3$, when one of the two inputs is a deterministic function of the feedback and the other input~\cite{elgamalmim}. In general, the capacity of the memoryless Coop2MAC is unknown~\cite{elgamalmim}.

The general Coop2MAC model allows the sources to operate in {\em full-duplex} mode (FD), i.e., to simultaneously send and receive. In practical systems however a node might either send or receive at any given time, but not both. In this case we say that the nodes operate in {\em half-duplex} mode (HD). In this work we consider both the FD and HD Gaussian Coop2MAC. We remark that there is no need to develop a separate theory for memoryless HD networks since the HD constraints can be incorporated into the memoryless FD framework as outlined in~\cite{kramerallerton}. In particular, for HD channels we slightly modify the model definition as follows: we let the channel input of user~$i\in\{1,2\}$ be the pair $(X_i,S_i)$, where as before $X_i\in \mathcal{X}_i$ and where the {\em state} $S_i\in\{0,1\}$ denotes whether the node is in receive-mode ($S_i=0$) or in transmit-mode ($S_i=1$). In other words, the HD channel is still memoryless but it is now specified by the four transitions probabilities, one for each possible pair $(S_1,S_2)\in\{0,1\}^2$.

\subsection{The Gaussian Coop2MAC}
In this work we focus on the Gaussian Coop2MAC because of its practical relevance: in the uplink of future cellular networks it is envisaged that mobiles will cooperate in order to increase their transmission rate to a central base station or enlarge the cell coverage. The simplest model to capture this scenario is the single-antenna complex-valued Gaussian FD Coop2MAC subject to an average power constraint that has input/output relationship
\begin{align}
& \begin{bmatrix}
  Y_1 \\ Y_2 \\ Y_3 \\
  \end{bmatrix}
  = \mathbf{H}
  \begin{bmatrix}
  X_1 \\ X_2 \\
  \end{bmatrix}
  +
  \begin{bmatrix}
  Z_1 \\ Z_2 \\ Z_3 \\
  \end{bmatrix},
  \mathbf{H} = \begin{bmatrix}
  \star  & \mathsf{h}_{\rm 1} \\
  \mathsf{h}_{\rm 2} & \star  \\
  \mathsf{h}_{\rm max} & \mathsf{h}_{\rm min} \\ 
  \end{bmatrix},
\label{eq:full half}
\end{align}
where the channel gains are complex-valued and constant (and therefore known to all nodes),
where $\star$ denotes a channel gain that does not affect the channel capacity (because self interference can be removed),
and where without loss of generality we assume that $|\mathsf{h}_{\rm max}|\geq|\mathsf{h}_{\rm min}|$ and refer to user~1 as the {\em strong user} (i.e., the source with the strongest link to the destination) and to user~2 as the {\em weak user}.
Without loss of generality, the inputs are subject to a unitary power constraint, i.e., $\mathbb{E}[ |X_i|^2 ] \leq 1$ for $i\in\{1,2\}$, and the jointly Gaussian noises are assumed to have zero mean and unit variance. In particular, but not without loss of generality, we assume that the noises are independent.

The HD channel is defined similarly to the FD one in~\eqref{eq:full half}.
The difference is that the channel matrix $\mathbf{H}$ becomes
\begin{align}
  \mathbf{H} =
  \begin{bmatrix}
  1-S_1 & 0   & 0   \\
  0   & 1-S_2 & 0   \\
  0   & 0     & 1   \\
  \end{bmatrix}
  \begin{bmatrix}
  \star  & \mathsf{h}_{\rm 1} \\
  \mathsf{h}_{\rm 2} & \star  \\
  \mathsf{h}_{\rm max} & \mathsf{h}_{\rm min} \\ 
  \end{bmatrix}
  \begin{bmatrix}
  S_1 & 0   \\
  0   & S_2 \\
  \end{bmatrix},
\label{eq:awgn half}
\end{align}
where $S_1$ and $S_2$ are binary-valued random variables representing the state of user~1 and user~2, respectively. 

\subsection{Generalized Degrees-of-Freedom Region (gDoF) and Capacity to within a Constant Gap}
The gDoF region is defined as follows~\cite{avestimer}.
For $\mathsf{SNR}>1$ parameterize the channel gains as
\[
|h_{i}|^2 := \mathsf{SNR}^{\beta_{i}}, \ \beta_{i}\geq 0, \ i\in\{{\rm 1},{\rm 2},{\rm max},{\rm min}\},
\]
and the rates as 
\[
r_i :=  \frac{R_i}{\log(1+\mathsf{SNR})}.
\]
The gDoF region is the set of all achievable pairs $(r_1,r_2)$ in the limit of $\mathsf{SNR}\to+\infty$.
The gDoF region is an asymptotically exact characterization of capacity at high $\mathsf{SNR}$.
At finite $\mathsf{SNR}$ the {\em capacity to within a constant gap} gives an approximate characterization of the capacity region.
The capacity is said to be known to within $\mathsf{b}$~bits if we can show an inner bound region  $\mathcal{I}$ and an outer bound region $\mathcal{O}$ such that 
$(R_1,R_2)\in{\rm ConvexClosure}[\mathcal{I}] \Longrightarrow (R_1+\mathsf{b},R_2+\mathsf{b})\not\in\mathcal{O}$.

\subsection{Past Work}
The study of the Coop2MAC was initiated in~\cite{willemsMACGF}, which prosed an achievable rate region based on partial-decode-forward and backward decoding; this region that is still the largest known to date. In Gaussian noise, Sendonaris et at~\cite{Sendonaris} studied the region of~\cite{willemsMACGF} for the FD case and proposed practical implementations for CDMA systems. For the fading Coop2MAC, power allocation schemes for both the ergodic and outage cases have been extensively studied; we will not revise them here for sake of space and because this work focuses on the static case. For the HD case, work such as~\cite{DavidsonSP,MaiVuISIT11} and references therein proposed inner and outer bound regions and numerically showed that they are not too far from one another; in this line of work, the optimization involved in order to find the largest achievable region is done numerically; although not specifically mentioned, these inner bound regions can be obtained from~\cite{willemsMACGF} by using the formalism of~\cite{kramerallerton}. As opposed to previous work, here we focus (a) on showing the asymptotic optimality of certain simple achievable schemes at high SNR and (b) on proving that the proposed schemes are optimality to within a constant gap for any channel parameter and at any SNR, in the spirit of~\cite{avestimer}. Our results can thus be considered as a step towards determining the capacity of the Gaussian Coop2MAC, both FD and HD, which to date is an open problem.

\subsection{Paper Organization}
The rest of the paper is organized as follows: Section~\ref{sec:main} summarizes the main results of the paper, Sections~\ref{sec:FD} and~\ref{sec:HD} contain the details of the proof for the FD and HD case, respectively, and Section~\ref{sec:conc} concludes the paper.

\section{Main Contributions}\label{sec:main}

\begin{figure*}
\centering
\subfigure[]{%
\includegraphics[width=0.45\textwidth]{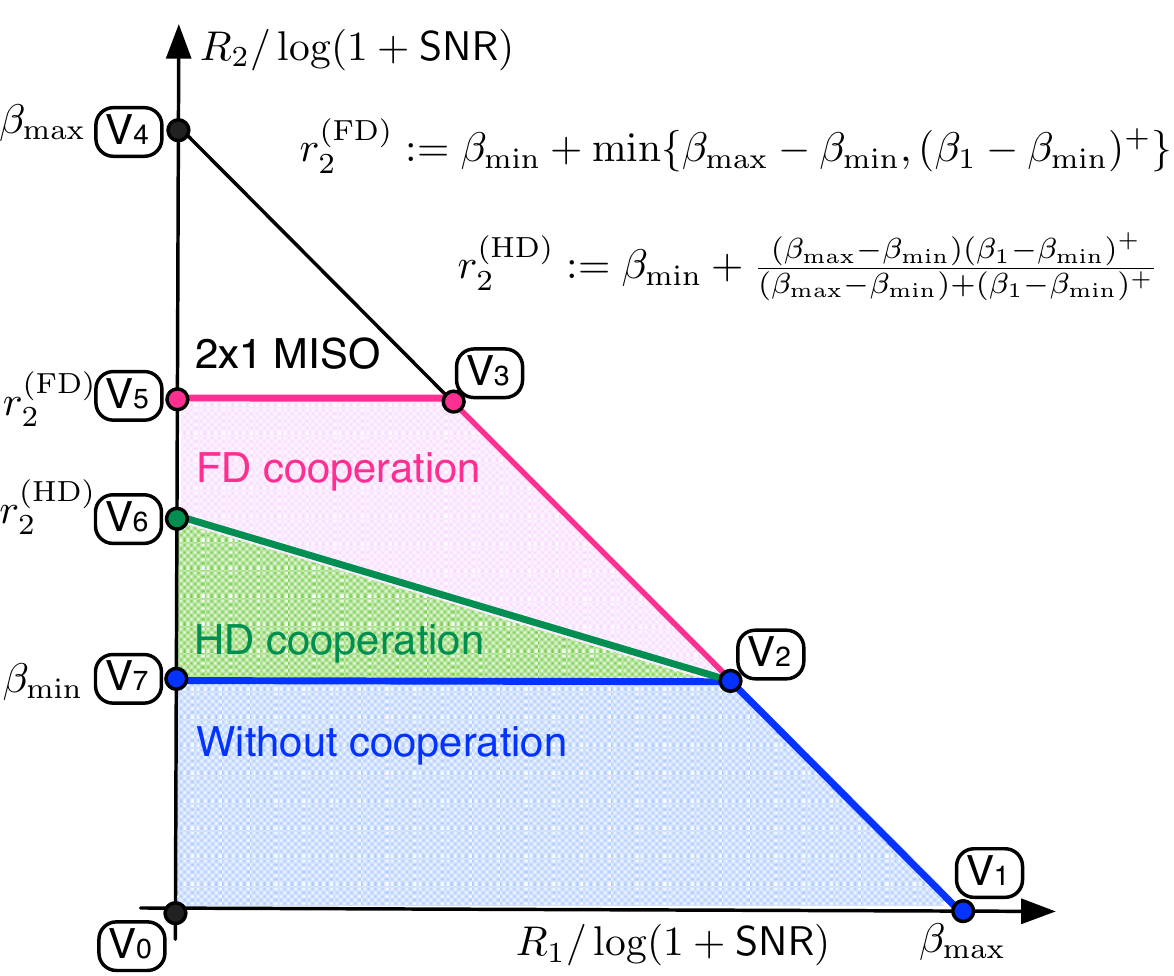}
\label{fig:capreg}
}%
\hspace*{1cm}
\subfigure[]{%
\includegraphics[width=0.45\textwidth]{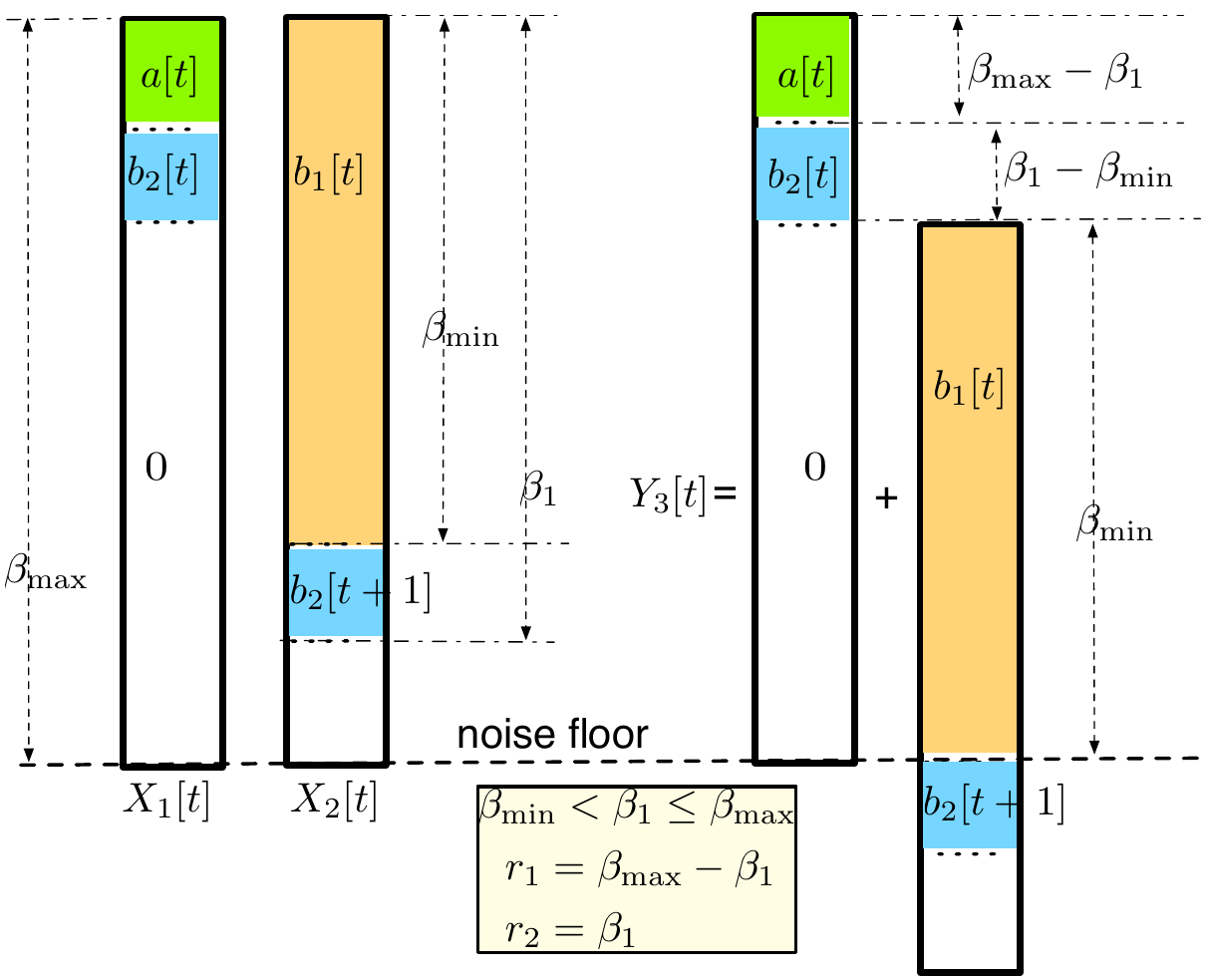}
\label{fig:achlindetch}
}%
\vspace*{-0.3cm}
\caption{%
(a) The gDoF region of the Gaussian Coop2MAC.
(b) An achievable scheme for the LDA.
}
\vspace*{-0.3cm}
%
%
\begin{align}
\hline 
&(R_1,R_2)=(R_1^{\rm(HD-RC)},R_2^{\rm(HD-RC)}) := \left(0,
\log(1+|\mathsf{h}_{\rm min}|^2) - \log(2) + \frac{c_{\rm max} \ c_{\rm 1}}{c_{\rm max}+c_{\rm 1}}
\right), \ \text{($c_{\rm max}$ and $c_{\rm 1}$ defined in~\eqref{eq:GD CS d})};
\label{eq:RC HD martina}
\\
&(R_1,R_2)=(R_1^{\rm(no-coop)},R_2^{\rm(no-coop)}) := \left(c_{\rm max}, \
\log\left(1+|\mathsf{h}_{\rm min}|^2\right)
\right), \ \text{($c_{\rm max}$ defined in~\eqref{eq:GD CS d})};
\label{eq:no coop}
\\&
\mathcal{C}^{\rm(HD)}
\subseteq
\bigcup
\left\{\begin{array}{l}
0\leq R_1 \leq I(X_1,S_1; Y_3, Y_2|X_2,S_2) \\
0\leq R_2 \leq I(X_2,S_2; Y_3, Y_1|X_1,S_1) \\
R_1+R_2   \leq I(X_1,S_1,X_2,S_2; Y_3)
\end{array}\right\}
\stackrel{\rm(a)}{\subseteq}
\bigcup_{\{\gamma_{ij}\}}\left\{\begin{array}{l}
0\leq R_1 \leq H^{(1)} +\sum_{(i,j)\in\{0,1\}^2} \gamma_{ij} I_{ij}^{(1)} \\
0\leq R_2 \leq H^{(2)} +\sum_{(i,j)\in\{0,1\}^2} \gamma_{ij} I_{ij}^{(2)} \\
R_1+R_2   \leq H^{(12)}+\sum_{(i,j)\in\{0,1\}^2} \gamma_{ij} I_{ij}^{(12)}
\end{array}\right\}
\label{eq:GD CS a}
\\&
\stackrel{\rm(b)}{\subseteq}
\bigcup_{\{\gamma_{ij}\}}\left\{\begin{array}{rl}
0\leq R_1 &\leq H^{(1)} + \gamma_{10}\log\frac{1}{\gamma_{10}}+ \gamma_{11}\log\frac{1}{\gamma_{11}} 
          + \gamma_{10}\log(1+|\mathsf{h}_{\rm 2}|^2+|\mathsf{h}_{\rm max}|^2) + \gamma_{11}\log(1+|\mathsf{h}_{\rm max}|^2) \\
0\leq R_2 &\leq H^{(2)} + \gamma_{01}\log\frac{1}{\gamma_{01}}+ \gamma_{11}\log\frac{1}{\gamma_{11}}
          + \gamma_{01}\log(1+|\mathsf{h}_{\rm 1}|^2+|\mathsf{h}_{\rm min}|^2) 
          +\gamma_{11}\log(1+|\mathsf{h}_{\rm min}|^2) \\
R_1+R_2   &\leq H^{(12)}+ \gamma_{10}\log\frac{1}{\gamma_{10}}
                       +\gamma_{01}\log\frac{1}{\gamma_{01}}
                       +\gamma_{11}\log\frac{1}{\gamma_{11}} +\gamma_{11}\log(2)\\
          &+ \gamma_{01}\log(1+|\mathsf{h}_{\rm min}|^2) + \gamma_{10}\log(1+|\mathsf{h}_{\rm max}|^2) + \gamma_{11}\log(1+|\mathsf{h}_{\rm max}|^2+|\mathsf{h}_{\rm min}|^2) \\
\end{array}\right\}
\label{eq:GD CS b}
\\&
\stackrel{\rm(c)}{\subseteq}
\bigcup_{\gamma\in[0,1]}\left\{\begin{array}{l}
0\leq R_1 ,\
0\leq R_2 \leq v^{(2)}  + \gamma\log(1+|\mathsf{h}_{\rm 1}|^2+|\mathsf{h}_{\rm min}|^2) + (1-\gamma)\log(1+|\mathsf{h}_{\rm min}|^2) \\
R_1+R_2   \leq v^{(12)} + \gamma\log(1+|\mathsf{h}_{\rm min}|^2) + (1-\gamma)\log(1+|\mathsf{h}_{\rm max}|^2+|\mathsf{h}_{\rm min}|^2) \\          
\end{array}\right\}
\label{eq:GD CS c}
\\&
\stackrel{\rm(d)}{\subseteq}
\mathcal{O}^{\rm(HD)} :=
\bigcup_{\gamma\in[0,1]}\left\{\begin{array}{l}
0\leq R_1 ,\
0\leq R_2 \leq v + \gamma    c_1, \\
R_1+R_2   \leq v + (1-\gamma)c_{\rm max} ,    
\end{array}\right\},
\left.\begin{array}{l}
v           := v^{(12)}+ \log(1+|\mathsf{h}_{\rm min}|^2), \ v^{(12)} := 3.8218~{\rm bits}, \\
c_1         := \log\left(1+\frac{|\mathsf{h}_{\rm 1}|^2  }{1+|\mathsf{h}_{\rm min}|^2}\right), \
c_{\rm max} := \log\left(1+\frac{|\mathsf{h}_{\rm max}|^2}{1+|\mathsf{h}_{\rm min}|^2}\right)\\ 
\end{array}\right..
\label{eq:GD CS d}
\\
\hline
\nonumber
\end{align}
\vspace*{-1.5cm}
\end{figure*}

Fig.\ref{fig:capreg} shows the gDoF regions for the Gaussian Coop2MAC, both FD and HD, for fixed $\mathsf{SNR}$-exponents $(\beta_{\rm 1},\beta_{\rm min},\beta_{\rm max})$.
A trivial achievable region can be obtained by ignoring the received generalized feedback signal at the sources, thereby obtaining a classical non-cooperative MAC whose capacity is 
\begin{align*}
\mathcal{C}^{\rm(no-coop)}
=
\left\{\begin{array}{l}
0\leq R_1 \leq \log(1+|\mathsf{h}_{\rm max}|^2) \\
0\leq R_2 \leq \log(1+|\mathsf{h}_{\rm min}|^2) \\
  R_1+R_2 \leq \log(1+|\mathsf{h}_{\rm max}|^2+|\mathsf{h}_{\rm min}|^2) 
\end{array}\right\}
\\
\Longleftrightarrow
\mathcal{C}^{\rm(no-coop)}_{\rm gDoF}
=\left\{\begin{array}{l}
0\leq r_1 ,\
0\leq r_2 \leq \beta_{\rm min} \\
  r_1+r_2 \leq \beta_{\rm max} 
\end{array}\right\},
\end{align*}
where the corner points of $\mathcal{C}^{\rm(no-coop)}_{\rm gDoF}$ are ${\rm V_0, V_1, V_2, V_7}$ in Fig.\ref{fig:capreg}; 
a trivial outer bound region can be obtained by letting the two sources exchange their messages ahead of transmission, thereby obtaining a $2\times1$ MISO channel, which we shall refer to as ``ideal cooperation'', whose capacity is 
\begin{align*}
\mathcal{C}^{\rm(ideal-coop)}
=
\left\{\begin{array}{l}
0\leq R_1, \ 0\leq R_2 \\
R_1+R_2 \leq \log(1+(|\mathsf{h}_{\rm max}|+|\mathsf{h}_{\rm min}|)^2) \\
\end{array}\right\}
\\
\Longleftrightarrow
\mathcal{C}^{\rm(ideal-coop)}_{\rm gDoF}
=
\left\{\begin{array}{l}
0\leq r_1, \ 0\leq r_2 \\
  r_1+r_2 \leq \beta_{\rm max} 
\end{array}\right\},
\end{align*}
where the corner points of $\mathcal{C}^{\rm(ideal-coop)}_{\rm gDoF}$ are ${\rm V_0, V_1, V_4}$ in Fig.\ref{fig:capreg}.

From these trivial bounds we immediately see that cooperation benefits only the weak user, in the sense that the weak user can achieve rates strictly above its maximum non-cooperative rate  only if the strong user reduces its rate below  its minimum non-cooperative rate  (i.e., move away from the point ${\rm V_2}$ in Fig.\ref{fig:capreg}).  Hence, with cooperation the rate of the weak user can have an unbounded improvement due to the possibility of {\em routing} its message through the strong user.  The results of this paper agree with this observation. 
In particular we shall show:

{\bf Theorem 1} (proved in Section~\ref{sec:FD}):
With FD cooperation the gDoF region has corner points ${\rm V_0, V_1, V_3, V_5}$ in Fig.\ref{fig:capreg}.

\begin{itemize}
\item
Point ${\rm V_1}$: achieved with no-cooperation.

\item
Point ${\rm V_3}$: achieved with the scheme in Fig.~\ref{fig:achlindetch}, which does not involve any backward or joint decoding.

\item
Point ${\rm V_5}$: achieved when the strong user acts as a pure relay for the weak user.
The capacity of such a FD Gaussian relay channel is known to within 1~bit~\cite{avestimer}.

\item
FD cooperation is equivalent to ideal cooperation gDoF-wise, i.e., the point ${\rm V_3}$ coincides with ${\rm V_4}$, for $\beta_{\rm 1}\geq \beta_{\rm max}$.

\item
FD cooperation reduces to no-cooperation gDoF-wise, i.e., the point ${\rm V_3}$ coincides with ${\rm V_2}$, for $\beta_{\rm 1}\leq \beta_{\rm min}$.

\item
For FD the capacity region can be established to within~\gapFD~bits with  unilateral cooperation.
\hfill $\blacksquare$
\end{itemize}

{\bf Theorem 2} (proved in Section~\ref{sec:HD}):
With HD cooperation the gDoF region has corner points ${\rm V_0, V_1, V_2, V_6}$ in Fig.\ref{fig:capreg}.

\begin{itemize}
\item
Points ${\rm V_1}$ and ${\rm V_2}$: achieved with no-cooperation.

\item
Point ${\rm V_6}$: achieved when the strong user acts as a pure relay for the weak user.
The capacity of such a HD Gaussian relay channel is known to within 3~bits~\cite{martinaICCsingleHDRC}.

\item
HD cooperation reduces to no-cooperation in terms of gDoF, i.e., the point ${\rm V_6}$ coincides with ${\rm V_7}$, for $\beta_{\rm 1}\leq \beta_{\rm min}$. 

\item
HD cooperation tends to ideal cooperation in terms of gDoF, i.e., the point ${\rm V_6}$ tends to ${\rm V_4}$, for $\beta_{\rm 1}\to +\infty$. 

\item
For HD the capacity can be established to within~\gapHD~bits by time sharing between the case where the two users do not cooperate and the case where the strong user acts as a pure relay for the weak user.  Unilateral cooperation suffices to achieve capacity to within a constant gap. 
\hfill $\blacksquare$
\end{itemize}

\section{The Full-Duplex Case}\label{sec:FD}

\subsection{Cut-set Upper Bound}
For a FD Coop2MAC with independent noises the cut-set upper bound gives $\forall \mathcal{S}\subseteq\{1,2\}\backslash\emptyset$, $\mathcal{S}^c = \{1,2\}\backslash\mathcal{S}$,  the bound $R_{\mathcal{S}} \leq I(X_{\mathcal{S}}; Y_3, Y_{\mathcal{S}^c}|X_{\mathcal{S}^c})$~\cite{elgamalmim}; by the `Gaussian maximizes entropy' principle~\cite{elgamalmim}, for all $\rho:=\mathbb{E}[X_1 X_2^*]$ such that $|\rho|\leq 1$, we have
\begin{align*}
R_1 &\leq 
\log(1+(|\mathsf{h}_{\rm max}|^2+|\mathsf{h}_{\rm 2}|^2)(1-|\rho|^2)), 
\\R_2 &\leq 
\log(1+(|\mathsf{h}_{\rm min}|^2+|\mathsf{h}_{\rm 1}|^2)(1-|\rho|^2)), 
\\R_1+R_2 &\leq 
\log(1+|\mathsf{h}_{\rm max}|^2+|\mathsf{h}_{\rm min}|^2 + 2 \mathrm{Re}\{\rho \mathsf{h}_{\rm max} \mathsf{h}_{\rm min}^*\}).
\end{align*}
By maximizing each bound in $\rho$ we find $\mathcal{C}^{\rm(FD)}\subseteq\mathcal{O}^{\rm(FD)}$ with
\begin{align*}
\mathcal{O}^{\rm(FD)} =
\left\{\begin{array}{l}
0\leq  R_1  ,\
0\leq  R_2 \leq \log(1+|\mathsf{h}_{\rm 1}|^2+|\mathsf{h}_{\rm min}|^2) \\
   R_1+R_2 \leq \log(1+(|\mathsf{h}_{\rm max}|+|\mathsf{h}_{\rm min}|)^2) 
\end{array}\right.
\\\Longleftrightarrow
\mathcal{O}^{\rm(FD)}_{\rm gDoF}
\subseteq
\left\{\begin{array}{l}
0\leq r_1 ,\
0\leq r_2 \leq \max\{\beta_{\rm min},\beta_{\rm 1}\} \\
  r_1+r_2 \leq \beta_{\rm max} 
\end{array}\right..
\end{align*}
From this upper bound on the gDoF region we see that cooperation benefits the weaker user only if $\beta_{\rm min} < \beta_{\rm 1} \leq \beta_{\rm max}$, in the sense that if $\beta_{\rm 1} \leq \beta_{\rm min}$ then $\mathcal{O}^{\rm(FD)}_{\rm gDoF}=\mathcal{C}^{\rm(no-coop)}_{\rm gDoF}$  (i.e., no point to cooperate if the inter-source channel is too weak), while if $\beta_{\rm 1} > \beta_{\rm max}$ then $\mathcal{O}^{\rm(FD)}_{\rm gDoF}=\mathcal{C}^{\rm(ideal-coop)}_{\rm gDoF}$ (i.e., no point to increase $\beta_{\rm 1}$ beyond $\beta_{\rm max}$).  Note also that $\beta_{\rm 2}$ (the channel gain from the weak user to the strong user) does not appear in the gDoF outer bound region.

\subsection{Achievability}
We next show the achievability of the corner points of the cut-set upper bound $\mathcal{O}^{\rm(FD)}$  to within a constant gap. 
The corner point with $R_2=0$ (equivalent to ${\rm V_1}$ in Fig.\ref{fig:capreg}) can be achieved to within 2~bits without cooperation since
\begin{align*}
\max_{(R_1,R_2)\in \mathcal{O}^{\rm(FD)}} \{ R_1 \} 
= \log(1+(|\mathsf{h}_{\rm max}|+|\mathsf{h}_{\rm min}|)^2)
\\
\leq \log(1+4|\mathsf{h}_{\rm max}|^2)
\leq \max_{(R_1,R_2)\in \mathcal{C}^{\rm(no-coop)}} \{ R_1 \} + \log(4),
\end{align*}
and the corner point with $R_1=0$ (equivalent to ${\rm V_5}$ in Fig.\ref{fig:capreg}) can be achieved to within 1~bit with either partial-decode-forward or compress-forward relaying~\cite{avestimer}.
The remaining corner point (equivalent to ${\rm V_3}$ in Fig.\ref{fig:capreg}) has coordinates
\begin{align*}
  &(R_1^{\prime\prime},R_2^{\prime\prime})=(\log(1+(|\mathsf{h}_{\rm max}|+|\mathsf{h}_{\rm min}|)^2) -t_2, \ t_2), 
\\& t_2 := \log(1+\min\{|\mathsf{h}_{\rm 1}|^2+|\mathsf{h}_{\rm min}|^2,(|\mathsf{h}_{\rm max}|+|\mathsf{h}_{\rm min}|)^2\}). 
\end{align*}
To show achievability of $(R_1^{\prime\prime},R_2^{\prime\prime})$ we distinguish three regimes for the cooperation gain $\mathsf{h}_{\rm 1}$:
%
\\{\em Regime 1:}
$|\mathsf{h}_{\rm 1}|^2 \leq |\mathsf{h}_{\rm min}|^2$ (equivalent to $\beta_{\rm 1} \leq \beta_{\rm min}$, that is, ${\rm V_5 = V_7 }$ and ${\rm V_3 = V_2 }$ in Fig.\ref{fig:capreg})
in which case $\min\{|\mathsf{h}_{\rm 1}|^2+|\mathsf{h}_{\rm min}|^2,(|\mathsf{h}_{\rm max}|+|\mathsf{h}_{\rm min}|)^2\}=|\mathsf{h}_{\rm 1}|^2+|\mathsf{h}_{\rm min}|^2 \in[|\mathsf{h}_{\rm min}|^2 , 2 |\mathsf{h}_{\rm min}|^2 ]$. The corner point can be achieved to within 1~bit without cooperation since
\begin{align*}
&R_2^{\prime\prime}=t_2 
 \leq \log(1+2|\mathsf{h}_{\rm min}|^2) 
 \leq \log(1+|\mathsf{h}_{\rm min}|^2) + \log(2),
\\ &\text{and} \quad
  R_1^{\prime\prime} = \log(1+(|\mathsf{h}_{\rm max}|+|\mathsf{h}_{\rm min}|)^2) -t_2
\\&\leq 
\log(1+|\mathsf{h}_{\rm max}|^2+|\mathsf{h}_{\rm min}|^2)+\log(2)-\log(1+|\mathsf{h}_{\rm min}|^2).
\end{align*}
%

\medskip
{\em Regime 2:}
$|\mathsf{h}_{\rm 1}|^2 > |\mathsf{h}_{\rm max}|^2+2 |\mathsf{h}_{\rm min}| |\mathsf{h}_{\rm max}|$  (equivalent to $\beta_{\rm 1} > \beta_{\rm max}$, that is, ${\rm V_5 = V_3 = V_4}$ in Fig.\ref{fig:capreg}) in which case $\min\{|\mathsf{h}_{\rm 1}|^2+|\mathsf{h}_{\rm min}|^2,(|\mathsf{h}_{\rm max}|+|\mathsf{h}_{\rm min}|)^2\}=(|\mathsf{h}_{\rm max}|+|\mathsf{h}_{\rm min}|)^2$. In this case the corner point has $R_1^{\prime\prime}=0$ and the rate $R_2^{\prime\prime}$ can be achieved to within 1~bit with either partial-decode-forward or compress-forward relaying~\cite{avestimer}.
%

\medskip
{\em Regime 3:}
$|\mathsf{h}_{\rm min}|^2 < |\mathsf{h}_{\rm 1}|^2 \leq  |\mathsf{h}_{\rm max}|^2+2|\mathsf{h}_{\rm min}| |\mathsf{h}_{\rm max}|$  (equivalent to $\beta_{\rm min} < \beta_{\rm 1} \leq \beta_{\rm max}$  in Fig.\ref{fig:capreg}) in which case $\min\{|\mathsf{h}_{\rm 1}|^2+|\mathsf{h}_{\rm min}|^2,(|\mathsf{h}_{\rm max}|+|\mathsf{h}_{\rm min}|)^2\}=|\mathsf{h}_{\rm 1}|^2+|\mathsf{h}_{\rm min}|^2$.  Achievability in this case requires a more sophisticated coding scheme, which we shall design next based on the insights we will gain from the {Linear Deterministic Approximation} (LDA) of the Gaussian noise channel at high $\mathsf{SNR}$~\cite{avestimer}. The LDA has input/output relationship
\[
Y_1 = \mathbf{S}^{n-\beta_{\rm 1}} X_2, \ 
Y_3 = \mathbf{S}^{n-\beta_{\rm max}} X_1 + \mathbf{S}^{n-\beta_{\rm min}} X_2,
\]
for some integers $\beta_{\rm max},\beta_{\rm min},\beta_{\rm 1}$, where the inputs and outputs are binary vectors of length $n:=\max\{\beta_{\rm max},\beta_{\rm min},\beta_{\rm 1}\}$, $\mathbf{S}$ is the $n\times n$ shift matrix~\cite{avestimer} and the additions are bit-wise on GF(2). 
Consider the scheme in Fig.\ref{fig:achlindetch}: 
signal $a$ is from user~1 and signal $(b_1,b_2)$ from user~2;
at time $t$ user~2 sends $\beta_{\rm min}$~bits directly to the destination through $b_1[t]$
and $\beta_{\rm 1}-\beta_{\rm min}$~bits to user~1 through $b_2[t+1]$;
the signal $b_2[t+1]$ appears below the noise floor of the destination
(who can only observe $\beta_{\rm min}$~bits of $X_2[t]$);
user~1 (who receives $\beta_{\rm 1}$~bits of $X_2[t]$) decodes both $b_1[t]$ and $b_2[t+1]$
and will forward $b_2[t+1]$ to the destination in the time slot $t+1$ on behalf of user~2;
at time $t$ user~1 sends $\beta_{\rm max}-\beta_{\rm 1}$~bits directly to the destination through $a[t]$
and $\beta_{\rm 1}-\beta_{\rm min}$~bits through $b_2[t]$;
at time $t$ the destination first decodes $a[t]$ achieving rate $r_1=\beta_{\rm max}-\beta_{\rm 1}$,
and then decodes $(b_1[t],b_2[t]$) achieving rate $r_2=(\beta_{\rm 1}-\beta_{\rm min})+(\beta_{\rm min})=\beta_{\rm 1}$.
Note that the weak user employs block Markov coding to convey information to the strong user but neither the destination nor the strong user use backward decoding~\cite{willemsMACGF}, i.e., the decoding incurs no delay.

We are now ready to show achievability to within a constant gap.
{Note that achievability could be shown by using Willem's coding scheme~\cite{willemsMACGF}, which involves block Markov coding and backward joint decoding. Instead, inspired by the scheme in Fig.\ref{fig:achlindetch} we describe a scheme that does not use backward decoding, which might be more relevant in practice because of its simplicity and because it does not incur in any delay.}
We start by assuming $1< |\mathsf{h}_{\rm min}|^2 = \min\{|\mathsf{h}_{\rm max}|^2,|\mathsf{h}_{\rm min}|^2,|\mathsf{h}_{\rm 1}|^2\}$.
Let $X_{a[t]},X_{b_1[t]},X_{b_2[t]},X_{b_2[t+1]}$ be i.i.d. (independent and identically distributed) $\mathcal{N}(0,1)$ and let the transmit signals in slot $t$ be
\begin{align*}
   X_1[t] &= \sqrt{1- \delta_1} \ X_{a[t]}   + \sqrt{\delta_1} \ X_{b_2[t]}, 
\\ X_2[t] &= \sqrt{1- \delta_2} \ X_{b_1[t]} + \sqrt{\delta_2} \ X_{b_2[t+1]}, \ 
\end{align*}
with
\begin{align*}
\delta_1 := \frac{|\mathsf{h}_{\rm 1}|^2}{|\mathsf{h}_{\rm max}|^2+2|\mathsf{h}_{\rm min}| |\mathsf{h}_{\rm max}|}, \
\delta_2 := \frac{1}{|\mathsf{h}_{\rm min}|^2}.
\end{align*}
User~1 receives 
\begin{align*}
Y_1[t]=\mathsf{h}_{\rm 1} \sqrt{1-\delta_2} \ X_{b_1[t]} 
           + \mathsf{h}_{\rm 1} \sqrt{  \delta_2} \ X_{b_2[t+1]}  + Z_1[t]
\end{align*}
and first decodes $X_{b_1[t]}$ by treating $X_{b_2[t+1]}$ as noise and then $X_{b_2[t+1]}$; this is possible if
\begin{align*}
   R_{b_1} &\leq 
             \log\left( 1+ |\mathsf{h}_{\rm 1}|^2\right) - \log\left( 1+ \frac{|\mathsf{h}_{\rm 1}|^2}{|\mathsf{h}_{\rm min}|^2}\right),
\\ R_{b_2} &\leq \log\left( 1+ \frac{|\mathsf{h}_{\rm 1}|^2}{|\mathsf{h}_{\rm min}|^2}\right).
\end{align*}
The destination receives 
\begin{align*}
Y_3[t]
  &=\mathsf{h}_{\rm max} \sqrt{1- \delta_1} \ X_{a[t]} + \mathsf{h}_{\rm max} \sqrt{\delta_1} \ X_{b_2[t]}
\\&+ \sqrt{|\mathsf{h}_{\rm min}|^2-1} \ X_{b_1[t]} +  X_{b_2[t+1]}+ Z_3[t]
\end{align*}
and successively decodes $X_{a[t]}$, $X_{b_2[t]}$ and $X_{b_1[t]}$ treating $X_{b_2[t+1]}$ as noise; this is possible if
\begin{align*}
   R_{a}   &\leq 
             \log\left( 1+ |\mathsf{h}_{\rm max}|^2 +  |\mathsf{h}_{\rm min}|^2 \right) 
 \\& - \log\left( 1+|\mathsf{h}_{\rm min}|^2 + \frac{|\mathsf{h}_{\rm max}|^2|\mathsf{h}_{\rm 1}|^2}{|\mathsf{h}_{\rm max}|^2+2|\mathsf{h}_{\rm min}| |\mathsf{h}_{\rm max}|}\right),
\\ R_{b_2} &\leq 
            \log\left( 1+|\mathsf{h}_{\rm min}|^2 + \frac{|\mathsf{h}_{\rm max}|^2|\mathsf{h}_{\rm 1}|^2}{|\mathsf{h}_{\rm max}|^2+2|\mathsf{h}_{\rm min}| |\mathsf{h}_{\rm max}|}\right)
\\& - \log\left(1+ |\mathsf{h}_{\rm min}|^2\right),
\quad 
R_{b_1} \leq 
             \log\left( 1+|\mathsf{h}_{\rm min}|^2 \right) - \log(2).
\end{align*}
Hence we achieve $R_1^\prime := R_{a}$ and $R_2^\prime := R_{b_1}+R_{b_2}$ with
\begin{align*}
R_2^\prime
=\log\left( 1+|\mathsf{h}_{\rm min}|^2 + |\mathsf{h}_{\rm 1}|^2 \frac{|\mathsf{h}_{\rm max}|}{|\mathsf{h}_{\rm max}| +2|\mathsf{h}_{\rm min}|}\right) 
-\log(2)
\end{align*}
Next, using $ |\mathsf{h}_{\rm 1}|^2 \leq |\mathsf{h}_{\rm max}|(|\mathsf{h}_{\rm max}|+2|\mathsf{h}_{\rm min}|)$, 
for the rate of the weak user we have
\begin{align*}
  R_2^{\prime\prime} - R_2^\prime 
  &= \log\left(\frac{1+|\mathsf{h}_{\rm min}|^2+|\mathsf{h}_{\rm 1}|^2}{1+|\mathsf{h}_{\rm min}|^2 + |\mathsf{h}_{\rm 1}|^2 \frac{|\mathsf{h}_{\rm max}|}{|\mathsf{h}_{\rm max}| +2|\mathsf{h}_{\rm min}|}}\right)+\log(2)
\\&
\leq 
\log\left(\frac{1+(|\mathsf{h}_{\rm max}|+|\mathsf{h}_{\rm min}|)^2}{1+|\mathsf{h}_{\rm min}|^2 + |\mathsf{h}_{\rm max}|^2}\right)+\log(2) \leq \log(4),
\end{align*}
since $(x+y)^2 \leq 2(x^2+y^2)$ for any $(x,y)\in\mathbb{R}^2$,
and for the rate of the strong user we have
\begin{align*}
  R_1^{\prime\prime} - R_1^\prime
&
\leq \log \left(
\frac{1+(|\mathsf{h}_{\rm max}|+|\mathsf{h}_{\rm min}|)^2}{1+|\mathsf{h}_{\rm min}|^2 + |\mathsf{h}_{\rm max}|^2} \right) \\&+
\log \left( \frac{1+|\mathsf{h}_{\rm min}|^2 + |\mathsf{h}_{\rm 1}|^2 \frac{|\mathsf{h}_{\rm max}|}{|\mathsf{h}_{\rm max}|+2|\mathsf{h}_{\rm min}|}}{1+|\mathsf{h}_{\rm min}|^2 + |\mathsf{h}_{\rm 1}|^2}
\right)
\\&
\leq \log(2\cdot 1) = \log(2).
\end{align*}

For $|\mathsf{h}_{\rm min}|^2 \leq 1$ we can use the same scheme we just described but with $R_{b_1}=0$; one can easily see that the achievable rate for user~1 remains $R_1^\prime$ as before while for user~2 one has to modify the expression of $R_2^\prime$ by substituting $\log(1+|\mathsf{h}_{\rm min}|^2)$ in place of $\log(2)$; hence the rate $R_2^\prime$ found before is a lower bound on the weak user's achievable rate under the condition $|\mathsf{h}_{\rm min}|^2 \leq 1$; since the previous gap analysis did not make use of the assumption $|\mathsf{h}_{\rm min}|^2 > 1$, the gap remains valid also for $|\mathsf{h}_{\rm min}|^2 \leq 1$. 

This concludes the proof of Thm.1.

\section{The Half-Duplex Case}\label{sec:HD}

From the analysis of the Gaussian FD Coop2MAC we known that the FD cut-set upper bound can be achieved to within 1~bit without cooperation if $|\mathsf{h}_{\rm 1}|^2 \leq|\mathsf{h}_{\rm min}|^2$; since HD can not do better than FD, we conclude that a no-cooperation scheme is optimal to within 1~bit for the Gaussian  HD Coop2MAC when $|\mathsf{h}_{\rm 1}|^2 \leq|\mathsf{h}_{\rm min}|^2$. Therefore we next only study the case $|\mathsf{h}_{\rm 1}|^2 > |\mathsf{h}_{\rm min}|^2$. We shall first describe a simple achievable scheme based on time sharing and then show its optimality to within~\gapHD~bits.

\subsection{Achievability}
The capacity of the Gaussian HD Relay Channel (HD-RC) is known to within 3~bits~\cite{martinaICCsingleHDRC}.
Therefore an achievable region for the HD Coop2MAC is obtained by time sharing between the rate attained when the strong user acts as pure relay for the weak user (equivalent to point $\rm V_6$ Fig.~\ref{fig:capreg}), given in~\eqref{eq:RC HD martina} {at the top of the previous page}, and the rate achieved without cooperation (equivalent to point $\rm V_2$ Fig.~\ref{fig:capreg}), given in~\eqref{eq:no coop} {at the top of the previous page}.

\subsection{Cut-set Upper Bound}
Under the HD condition the cut-set upper bound gives the region in~\eqref{eq:GD CS d} {at the top of the previous page}, where the different inclusions can be proved as follows.
For the inequality in~\eqref{eq:GD CS a} {at the top of the previous page}:
for $(i,j)\in\{0,1\}^2$, let
$\gamma_{ij}  := P[S_1=i,S_2=j]  \in[0,1]$ such that $\sum_{(i,j)\in\{0,1\}^2} \gamma_{ij}=1$,
and, conditioned on $(S_1=i,S_2=j)$,  let the input covariance matrix be
\begin{align*}
\begin{bmatrix}
P_{1,ij} & \rho_{ij} \sqrt{P_{1,ij} P_{2,ij}} \\
\rho_{ij}^* \sqrt{P_{1,ij} P_{2,ij}} & P_{1,ij} \\
\end{bmatrix} : |\rho_{ij}|\leq 1, \ 
\end{align*}
so that the power constraint can be expressed as
\begin{align}
  \sum_{(i,j)\in\{0,1\}^2} \gamma_{ij} P_{k,ij} \leq 1, \ k\in\{1,2\};
  \label{eq:power constraint}
\end{align}
for $\mathcal{S}\subseteq\{1,2\}\backslash\emptyset$, $\mathcal{S}^c = \{1,2\}\backslash\mathcal{S}$, let
\[
I(X_{\mathcal{S}}; Y, Y_{\mathcal{S}^c}|X_{\mathcal{S}^c},S_1=i,S_2=j) =: I_{ij}^{(\mathcal{S})},
\]
and, for discrete random variables,
\[
I(S_{\mathcal{S}} ; Y, Y_{\mathcal{S}^c}|X_{\mathcal{S}^c}, S_{\mathcal{S}^c})
\leq H(S_{\mathcal{S}}|S_{\mathcal{S}^c}) 
=: H^{(\mathcal{S})};
\]
then the inclusion in~\eqref{eq:GD CS a} follows from the above definitions and because 'Gaussian maximizes entropy'.
For the inequality in~\eqref{eq:GD CS b} {at the top of the previous page}:
rewrite the power constraint in~\eqref{eq:power constraint} as
\[
P_{k,ij} = \frac{\delta_{k,ij}}{\gamma_{ij}} : 
\delta_{k,ij}\in[0,1] \ {\rm and} \!\!\!
\sum_{(i,j)\in\{0,1\}^2} \delta_{ij} \leq 1, \ k\in\{1,2\};
\]  
then, $I_{00}^{(1)} = I_{01}^{(1)} = I_{00}^{(2)} = I_{10}^{(2)} = I_{00}^{(12)} = 0$ and
\begin{align*}
  I_{10}^{(1)} 
  &\leq \log(1+(|\mathsf{h}_{\rm 2}|^2+|\mathsf{h}_{\rm max}|^2)(1-|\rho_{10}|^2)P_{1,10})
\\&\leq \log(1+(|\mathsf{h}_{\rm 2}|^2+|\mathsf{h}_{\rm max}|^2) \ P_{1,10})
\\&= \log(\gamma_{10}+(|\mathsf{h}_{\rm 2}|^2+|\mathsf{h}_{\rm max}|^2) \ \delta_{1,10})- \log(\gamma_{10})
\\&\leq \log(1+|\mathsf{h}_{\rm 2}|^2+|\mathsf{h}_{\rm max}|^2)- \log(\gamma_{10})
\end{align*}
and similarly for $I_{11}^{(1)}, I_{01}^{(2)}, I_{11}^{(2)}, I_{01}^{(12)}, I_{10}^{(12)}, I_{11}^{(12)}$.
%
For the inequality in~\eqref{eq:GD CS c} {at the top of the previous page}:
we drop the constraint on $R_1$ and we bound
\begin{align*}
  &H^{(1)}
   + \gamma_{10} \log \frac{1}{\gamma_{10}}
   + \gamma_{11} \log \frac{1}{\gamma_{11}}
\leq v^{(1)} =: 2.0182~{\rm bits},
\end{align*}
with equality for $\gamma_{00}=\gamma_{01} = t^2\exp(1), \gamma_{10} = \gamma_{11} = t$
for some $t\in\mathbb{R}_+$ such that $\sum_{(i,j)\in\{0,1\}^2} \gamma_{ij}=1$;
similarly for the bound on $R_2$ we find 
\[
  H^{(2)}
   + \gamma_{01} \log \frac{1}{\gamma_{00}}
   + \gamma_{11} \log \frac{1}{\gamma_{11}}
\leq v^{(2)}=v^{(1)};
\]
with a similar reasoning
\begin{align*}
  &H^{(12)} 
   + \gamma_{10} \log \frac{1}{\gamma_{10}}
   + \gamma_{01} \log \frac{1}{\gamma_{01}}
   + \gamma_{11} \log \frac{1}{\gamma_{11}}
   + \gamma_{11} \log 2
\\& \leq v^{(12)} = 3.8218~{\rm bits},
\end{align*}
with equality for $\gamma_{00} = t^2\exp(1), \gamma_{01}=\gamma_{10} = t, \gamma_{11} = t\sqrt{2}$
for some $t\in\mathbb{R}_+$ such that $\sum_{(i,j)\in\{0,1\}^2} \gamma_{ij}=1$;
finally, we use $\log(1+|\mathsf{h}_{\rm max}|^2)\leq \log(1+|\mathsf{h}_{\rm max}|^2+|\mathsf{h}_{\rm min}|^2)$ and $v^{(2)}\leq v^{(12)}$.

In order to find the optimal $\gamma\in[0,1]$, representing the optimal fraction of time the strong user listens to the channel in order to help the weak user, for each point on the convex closure of the outer bound in~\eqref{eq:GD CS d}, {at the top of the previous page}, we rewrite $\mathcal{O}^{\rm(HD)}= \mathcal{O}^{\rm(HD \ 1)} \cup \mathcal{O}^{\rm(HD \ 2)}$ with
\begin{align*}
&
\mathcal{O}^{\rm(HD \ 1)} 
:=
\bigcup_{\gamma\in\left[0,\frac{c_{\rm max}}{c_{\rm max}+c_{\rm 1}}\right]}
\left\{\begin{array}{l}
    R_2 \leq v+\gamma    c_{\rm 1}  \\
R_1+R_2 \leq v+(1-\gamma)c_{\rm max}\\
\end{array}\right\}
\\&
\mathcal{O}^{\rm(HD \ 2)}
:=
\bigcup_{\gamma\in\left[\frac{c_{\rm max}}{c_{\rm max}+c_{\rm 1}},1\right]}
\left\{\begin{array}{l}
R_1+R_2 \leq v+(1-\gamma)c_{\rm max}\\
\end{array}\right\}
\\&\quad =\left\{\begin{array}{l}
R_1+R_2 \leq v+\frac{c_{\rm max}c_{\rm 1}}{c_{\rm max}+c_{\rm 1}}\\
\end{array}\right\}
\end{align*}
since $(1-\gamma)c_{\rm max} < \gamma c_{\rm 1}$ implies $\gamma > \frac{c_{\rm max}}{c_{\rm max}+c_{\rm 1}}$.
Next, for any $\mu\in[0,1]$ we solve $\max\{\mu R_1 + (1-\mu) R_2\}$ subject to $(R_1,R_2)\in\mathcal{O}^{\rm(HD)}$.
For $\mu\in[0,1/2)$ we have
\begin{align*}
\mathsf{p}(\mu)
  &:=
  \max_{(R_1,R_2)\in\mathcal{O}^{\rm(HD \ 1)}}\left\{ \mu R_1 + (1-\mu) R_2 \right\}
\\&= \max_{(R_1,R_2)\in\mathcal{O}^{\rm(HD \ 1)}}\left\{ \mu (R_1+R_2) + (1-2\mu) R_2 \right\}
\\&\leq (1-\mu)v
  +\max_{\gamma\in[0,\frac{c_{\rm max}}{c_{\rm max}+c_{\rm 1}}]}\left\{  
      \mu (1-\gamma) c_{\rm max} + (1-2\mu) \gamma c_{\rm 1} 
\right\}.
\end{align*}
Then, in the last optimization in the above formula, if $(1-2\mu)c_{\rm 1}-\mu c_{\rm max}\leq 0$ then the optimal is $\gamma=0$ (i.e., the strong user does not help the weak user)
otherwise the optimal is $\gamma=\frac{c_{\rm max}}{c_{\rm max}+c_{\rm 1}}$.
Please note that we found a closed-form expression for the optimal fraction of time the strong user listens to the channel, in the spirit of~\cite{martinaICCsingleHDRC}.
With the optimal $\gamma$ we have
\begin{align*}
  &\mathsf{p}(\mu)=(1-\mu)v^{(12)}+(1-\mu)R_2^{\rm(no-coop)} + \mu R_1^{\rm(no-coop)} 
\\&\quad \text{if} \ \frac{c_{\rm 1}}{c_{\rm max}+2c_{\rm 1}} \leq  \mu < \frac{1}{2} \ \text{(i.e., no-cooperation is optimal)},
\\
  &\mathsf{p}(\mu)=(1-\mu)(v^{(12)}+\log(2))+(1-\mu)R_2^{\rm(HD)}+\mu R_1^{\rm(HD)}
\\&\quad \text{if} \  0\leq \mu < \frac{c_{\rm 1}}{c_{\rm max}+2c_{\rm 1}} \ \text{(i.e., pure relaying is optimal)},
\end{align*}
which is exactly the achievable region based on time-sharing
described above (equivalent to the segment between ${\rm V_6}$ and ${\rm V_2}$ in Fig.~\ref{fig:capreg})
up to a gap of $(v^{(12)}+\log(2))$~=~\gapHD~bits.
For $\mu\in[1/2,1]$ we have
\begin{align*}
  &\max_{(R_1,R_2)\in\mathcal{O}^{\rm(HD \ 1)}}\left\{ \mu R_1 + (1-\mu) R_2 \right\}
\\&= \max_{(R_1,R_2)\in\mathcal{O}^{\rm(HD \ 1)}}\left\{ \mu (R_1+R_2) - (2\mu-1) R_2 \right\}
\\&\leq  \mu \big(v+c_{\rm max} \big),
\end{align*}
which is exactly the achievable region based on time-sharing
described above (equivalent to the segment between ${\rm V_1}$ and ${\rm V_2}$ in Fig.~\ref{fig:capreg})
up to a gap of $v^{(12)}<$~\gapHD~bits.

From the above characterization of the region $\mathcal{O}^{\rm(HD \ 1)}$ we immediately see that 
$\mathcal{O}^{\rm(HD \ 2)} \subseteq \mathcal{O}^{\rm(HD \ 1)}$
(i.e., with reference to  Fig.~\ref{fig:capreg}:
$\mathcal{O}^{\rm(HD \ 2)}$ is the equivalent to the region 
with corner points ${\rm V_0,V_6}$ and the one obtained by drawing a line at -45~degrees through ${\rm V_6}$, while $\mathcal{O}^{\rm(HD \ 1)}$ is the region with corner points ${\rm V_0,V_1,V_2,V_6}$).

This concludes the proof of Thm.2.

\section{Conclusions}\label{sec:conc}
This work characterized the degrees of freedom region and the capacity to within a constant gap for the two-user Gaussian Multiple Access Channel where the two users cooperates over a noisy channel. Both the full-duplex and the half-duplex case were investigated. In both cases unilateral cooperation and relatively simple achievable schemes were shown to suffice to achieve the cut-set upper bound to within a constant gap.

\section*{Acknowledgement}
The work of D.~Tuninetti was partially funded by NSF under award number 0643954;
the contents of this article are solely the responsibility of the author and
do not necessarily represent the official views of the NSF.
The work of Dr. D.~Tuninetti was possible thanks to the generous support of Telecom-ParisTech, Paris France, while the author was on a sabbatical leave at the same institution.

\bibliographystyle{IEEEbib}

\end{document}